# Transverse Rashba Effect and Unconventional Magnetocrystalline Anisotropy in Double-Gd-adsorbed Zigzag Graphene Nanoribbon

Weifeng Xie[a], Yu Song[b,†], Xu Zuo[a,c,d,e,*]

[a] *College of Electronic Information and Optical Engineering,*
*Nankai University, Tianjin, 300350, China;*
[b] *College of Physics and Electronic Information Engineering,*
*Neijiang Normal University, Neijiang, 641112, China;*
[c] *Key Laboratory of Photoelectronic Thin Film Devices and*
*Technology of Tianjin, Tianjin, 300071, China;*
[d] *Engineering Research Center of Thin Film Optoelectronics Technology, Ministry*
*of Education, Tianjin, 300071, China;*
[e] *Institute of Electronic Engineering,*
*China Academy of Engineering Physics, Mianyang, 621999, China.*
[†] *Previously with Microsystem and Terahertz Research Center,*
*China Academy of Engineering Physics, Chengdu,*
*610200, China; and Institute of Electronic Engineering, China*
*Academy of Engineering Physics, Mianyang, 621999, China.*
[*] *Corresponding author: xzuo@nankai.edu.cn*

## Abstract

The transverse Rashba effect is proposed and investigated by the first-principle calculations based on density functional theory in a quasi-one-dimensional antiferromagnet with a strong perpendicular magnetocrystalline anisotropy, which is materialized by the Gd-adsorbed graphene nanoribbon with a centric symmetry. The Rashba effect in this system is associated with the local dipole field transverse to and in the plane of the nanoribbon. That dipole field is induced by the off-center adsorption of the Gd adatom above the hex-carbon ring near the nanoribbon edges. The transverse Rashba effect at the two Gd adatoms enhances each other in the antiferromagnetic (AFM) ground state and cancels each other in the ferromagnetic (FM) meta-stable state, because of the centrosymmetric atomic structure. The transverse Rashba parameter is 1.51 eVÅ. This system shows a strong perpendicular magnetocrystalline anisotropy (MCA), which is 1.4 meV per Gd atom in the AFM state or 2.2 meV per Gd atom in the FM state. The origin of the perpendicular MCA is analyzed in $k$−space by filtering out the contribution of the transverse Rashba effect from the band structures perturbed by the spin-orbit coupling interactions. The first-order perturbation of the orbit and spin angular momentum coupling is the major source of the MCA, which is associated with the one-dimensionality of the system. The transverse Rashba effect and the strong perpendicular magnetization hosted simultaneously by the proposed AFM Gd-adsorbed graphene nanoribbon lock the up- (or down-) spin quantization direction to the forward (or backward) movement. This finding offers a magnetic approach to a high coherency spin propagation in one-dimensionality, and open a new door to manipulating spin transportation in graphene-based spintronics.

I. INTRODUCTION

Spintronics, one of the most promising technologies for next-generation information engineering[1], have been being investigated extensively. Spin in addition to charge degree of freedom in spintronics is employed for information storage, transmission, and processing, which may essentially reshape the electronics with higher density, faster speed, smaller die size, and lower power consumption. The Rashba effect, of which the Hamiltonian is written as, $\mathbf{H}_R = (\alpha_R/\hbar)(\boldsymbol{\sigma}\times\mathbf{p})\cdot\mathbf{e}$, is prospective for manipulating spin transportation in spintronics, as it couples electron spin $\boldsymbol{\sigma}$ to its own momentum $\mathbf{p}$ in electric field along direction $\mathbf{e}$[2–4]. The signature of the Rashba effect on band structure is the so-called spin-splitting, where the bands with opposite spins shift toward opposite directions in $k-$space. A giant Rashba spin splitting was recently discovered in bulk materials, such as BiTeI[5, 6], due to the non-centrosymmetric crystal structure and the strong spin-orbital coupling (SOC) interaction of heavy elements. It was also discovered in the interfaces, such as Bi/Ag(111)[7], where the inversion symmetry is broken by the interface.

Although it is generally recognized that breaking inversion symmetry is indispensable for the Rashba effect of nonmagnetic materials, the so-called R-2 type spin polarization that is generated by the Rashba effect was recently proposed in three-dimensional nonmagnetic crystals with inversion symmetry[8]. In fact, if a centrosymmetric crystal structure can be divided into two sectors associated with the inversion symmetry and there is a local dipole field in each sector, the Rashba effect will emerge and generate local spins opposite in direction in the two sectors. It should be noted that the local spins compensate each other and the total spin vanishes, as required by the inversion symmetry[9–15]. This implies that the Rashba effect may appear in antiferromagnets with inversion symmetry as an inverse effect of the R-2 type spin polarization in centrosymmetric nonmagnetic crystals, in the sense that the compensated spin polarization of antiferromagnet accommodates the Rashba effect.

The Rashba effect was also investigated in magnetic materials[16, 17]. It's worth noting that, in magnetic materials, although the states for opposite magnetization directions do not exist simultaneously in ferromagnets, the Rashba splitting in the SOC band structures is also equivalent to the Rashba-type spin splitting in nonmagnetic materials[16]. In the past few decades, the Rashba effect in magnetic materials was discovered in graphene layer on Ni(111) surface[18], and was predicted for the ultrathin ferromagnetic film on nonmagnetic heavy-metal layer[19], one- or two-dimensional space inversion symmetry broken antiferromagnet[20], multi-ferromagnetic perovskite[21], and ferromagnetic Gd-adsorbed graphene nanoribbon[22, 23]. In the interfaces composed of nonmagnetic and ferromagnetic layers, the electric field perpendicular to the interface plane breaks the space inversion symmetry and couples itself to the in-plane components of spin and momentum. This so-called perpendicular Rashba effect will vanish, when the magnetic easy axis is perpendicular to the interface plane. On the contrary, the Rashba effect of the Gd-adsorbed zigzag graphene nanoribbon (ZGNR) involves the local electric field transverse to the nanoribbon, and thus called the transverse Rashba effect. As the transverse electric field is coupled to the perpendicular spin component and the longitudinal crystalline momentum, it is possible to utilize this effect to manipulate spin transportation even if there is perpendicular magnetic anisotropy. Moreover, the transverse electric field is local in nature, since it is associated with the off-center adsorption on the edges of the nanoribbon[22]. It is thus possible to decorate the nanoribbon with more than one magnetic heavy-metal adatoms per unit cell to achieve the transverse Rashba effect in a more complicated spin structure, such as antiferromagnetism.

Magnetic crystalline anisotropy (MCA) is one of the most fundamental magnetic properties. Jointly with other parameters, it defines the magnetic behaviors of a magnet, such as the magnetic easy axis, the shape of hysteresis loop, and the stability of magnetization at finite temperature, which will further impact the design and implementation of magnetic devices. For example, a strong perpendicular MCA is preferred for the magnetic films in magnetic storage devices, such as hard disk, to boost the number of bits per unit area and enhance the stability against the external noises induced by thermal fluctuation, magnetic field, and electric current. Moreover, antiferromagnetism is preferred for the integrated spintronics to minimize the stray field[24–26], although the MCA of antiferromagnets was rarely investigated.

MCA is induced by the coupling between orbit and spin angular momentums ($LS-$coupling) inside magnetic atoms[27, 28]. An accurate first-principle calculation of MCA has been pursued from decades. The first-principle MCA is usually derived from the total energies calculated with the SOC perturbation for different magnetization directions[29, 30]. The force theorem is the essence of the MCA calculations, which requires that the spin-densities (up and down) are approximately invariant for different magnetization directions. However, the early calculations of MCA were suffered from the insufficient quality of the calculated band structures. In fact, the $LS$-coupling interaction perturbs the bands near the Fermi energy[28] and induces spikes to the $k$-space distribution of MCA for metals, which requires a $k$-mesh dense enough to achieve convergence. Several approaches such as state tracking method[31] and torque method[32, 33] were proposed to solve this problem with an acceptable computation cost at that time. MCA can also be calculated in a self-consistent way. This approach was applied to a free-standing Rh monolayer[34]. It was shown that the convergence and fluctuation of the MCA were influenced by not only the number of $k$-points sampled in the surface Brillouin zone but also the integration scheme. Electron correlation also influences the first-principle MCA, as it may modify the band structure. A LSDA+U calculation predicted both the correct easy axes and the magnitudes of MCA for Fe and Ni[35].

The first-principle MCA can be quantitatively analyzed by considering the $k$-space distribution of MCA, that is, the difference of the summation of all occupied Kohn-Sham eigenvalues at each $k$-point induced by rotating the magnetization direction. The first-order perturbation of contributes to the MCA at the $k$-points where the energy difference between the occupied and empty states is comparable to the $LS$-coupling constant ($\xi$) in the Hamiltonian $\boldsymbol{H}_{LS} = \xi \boldsymbol{L} \cdot \boldsymbol{S}$. This contribution is called surface pair coupling (SPC). It was estimated for two-dimensional materials that the SPC contribution is proportional to $\xi^3$, which is an order higher than the second-order perturbation that is proportional to $\xi^2$[31]. However, the SPC contribution can be the major one to MCA for some one-dimensional materials with a special band structure. Assume that the valence and conduction bands are parabolically contacted at the $\Gamma$-point in the one-dimensional Brillouin zone, in addition, with a perfect electron-hole symmetry for mathematical simplicity. It follows that the length of SPC region is $L_{SPC} \approx (2/\hbar)\sqrt{m^*\xi}$, where $m^*$ is the effective mass at $\Gamma$. The SPC contribution to MCA is then proportional to $\xi^{\frac{3}{2}}$, which is a half order lower than the second-order perturbation.

In addition, it was discovered by the first-principle calculations that both the Rashba effect and the $LS$-coupling interaction perturb the band structure of the Gd-adsorbed ZGNR when the magnetization is perpendicular to the nanoribbon. It was proposed that the perturbation due to the transverse Rashba effect can be extracted in $k$-space by simply taking the difference of the SOC perturbation calculated with the [001] and [00$\bar{1}$] magnetization directions, respectively, and that the resulting perturbation due to the transverse Rashba effect is an odd function in the one-dimensional Brillouin zone. In fact, the transverse Rashba effect is antisymmetric with respect to flipping the spin from the [001] direction to the [00$\bar{1}$] direction. Similarly, the $LS-$coupling perturbation can be extracted in $k-$space by taking the summation of the SOC perturbation calculated with the [001] and [00$\bar{1}$] magnetization directions, respectively, as it is symmetric with respective to flipping the spin direction.

In this work, an artificial structure of double-Gd-adsorbed ZGNR with a centric symmetry is proposed and investigated by the first-principle calculations based on density functional theory. It is shown that the structure simultaneously accommodates the antiferromagnetic ground state, the MCA perpendicular to the nanoribbon plane, and the Rashba effect by the local in-plane dipole field transverse to the nanoribbon. The dipole field is associated with the off-center adsorption of the Gd-adatoms on top of the hex-carbon rings near the edges. It couples the spin quantization direction defined by the perpendicular MCA to the crystalline momentum ($\hbar \mathbf{k}$). This enables the manipulation of spin transportation in a quasi one-dimensional antiferromagnet with perpendicular MCA via the transverse Rashba effect. The origin of the perpendicular MCA is analyzed by considering the $k-$space distribution extracted from the SOC perturbed band structures. It is shown that the first-order perturbation of $LS-$coupling between the valence band maximum and the conduction band minimum at the $\Gamma-$point is the major source of the perpendicular MCA. As a comparison, the same atomic structure in the ferromagnetic state is investigated. The transverse Rashba effect of the two Gd-adatoms, however, cancels each other in the ferromagnetic state. The MCA is perpendicular, and the $k-$space distribution of MCA shows several spikes associated with the metallic band structure of the ferromagnetic state. The first-order perturbation of the $LS-$coupling interaction is the major source of the MCA by directly and indirectly changing the occupation number of the states near the Fermi energy.

## II. APPROACH

First-principles calculations based on density functional theory are performed by using the projected augmented wave (PAW) [36] method as implemented in the Vienna $ab-initio$ simulation package (VASP)[37]. SOC is included in the calculations for both antiferro-magnetic (AFM) and ferromagnetic (FM) states to investigate the Rashba effect and the MCA. Exchange-correlation energy functional is treated in the Perdew-Buke-Ernzerhof parameterization of generalized gradient approximation (GGA-PBE)[38]. The valence electron configuration of Gd atom is chosen to be $4f^75d^16s^2$. The wave functions are expanded into a plane wave basis with a kinetic energy cutoff of 450 eV. It's necessary to induce the correction of Hubbard $U$ term (GGA+$U$)[39] to describe the strongly localized $f$ orbits of Gd atom, and $U_f$=8 eV is adopted[22, 40]. The unit cell (shown in the dotted box in FIG. 1 (a) and (b)) is isolated from its periodic images by a vacuum layer as thick as 15 Å along the transverse direction ($x$−axis, [100] direction) and that as thick as 12 Å along the perpendicular direction ($z$−axis, [001] direction). The nanoribbon is along the $y$−axis ([010] direction). The carbon dangling bonds at the edges are passivated by hydrogens. The first Brillouin zone (FBZ) is sampled by a 1×29×1 Monkhorst-Pack grid for the structure optimization, where the atomic positions are fully relaxed with the conjugate gradient algorithm until the force on each atom is less than 0.01 eV Å$^{-1}$ and the energy error is less than 1×10$^{-7}$ eV. In the band structure calculations from which the total energies are extracted, a 1×69×1 $k$−mesh is hired for the Brillouin zone integration. According to the force theorem[29, 30], considering the SOC interaction as a perturbation, the uniaxial MCA constants ($K_u$) for the AFM and FM states are calculated from the total energy difference between the $y$−axes ([010]) and $z$−axes ([001]) magnetization directions, namely $K_u=E_{[010]}-E_{[001]}$. In this work, the convergence of the calculated MCA versus the number of $k$−points has been tested firstly and plotted in FIG. S1 of Supplementary Information[41]. The dipole correction[42] is applied in all the calculations.

### III. RESULTS AND DISCUSSION

#### A. Stability of ZGNR decorated with two Gd atoms

The stability of the two Gd atoms adsorbed on ZGNR with index N=4 (2Gd-4ZGNR) system (FIG. 1 (a)) is investigated in AFM and FM states. The pure ZGNR with index N=4 (4ZGNR) where the carbon dangling bonds at the edges are passivated by hydrogens contains 32 carbons and 8 hydrogen atoms, which is fully relaxed before adsorbing two Gd atoms. The relaxed results show that the 4ZGNR exhibits an AFM semiconductor property which is in agreement with the previous reports[41, 43, 44]. Then the adsorption positions of the two Gd adatoms on 4ZGNR are considered to accommodate a centric symmetry, and the unit cell constitutes a coverage of two Gd adatoms per 32 carbon atoms, shown in FIG. 1 (a). Considering two magnetic states of 2Gd-4ZGNR. Firstly, for AFM state, before fully relaxation, two Gd adatoms with opposite magnetic moment adsorbed on the relaxed 4ZGNR with AFM ground state, one Gd atom places in directly above the hex-carbon ring near the edge and the other Gd atom places in directly below the hex-carbon ring near the other edge, and both Gd adatoms have an identical distance from point ′O′ (the center of the local edge hexacarbon ring, shown in FIG. 1 (c)). After fully relaxation, as shown in FIG. 1 (a) and (b), the 2Gd-4ZGNR is AFM state. Moreover, two Gd adatoms are prefer more deviation from the directly over the point ′O′ (FIG. 1 (c)), and have an off-center adsorption. It is consistent with the previous reports in which the near-hollow or hollow site is more favorable than the top or bridge site for Gd adsorbed on graphene nanoribbon or graphene[22, 45]. For FM state, before fully relaxation, two Gd adatoms with the same magnetic moment adsorbed on the relaxed 4ZGNR with the FM metastable state, and the adsorption locations of two Gd adatoms are the same as that for AFM state. After fully relaxation, the whole system is FM state which is higher than the AFM state by 45.383 meV in energy. Similarly, two Gd atoms also have an off-center adsorption. As a comparison, the structure as well as the electronic properties of the 2Gd-4ZGNR corresponding to a coverage of two Gd adatoms per 16 carbon atoms has also been explored, shown in FIG. S2 of Supplementary Information[41].

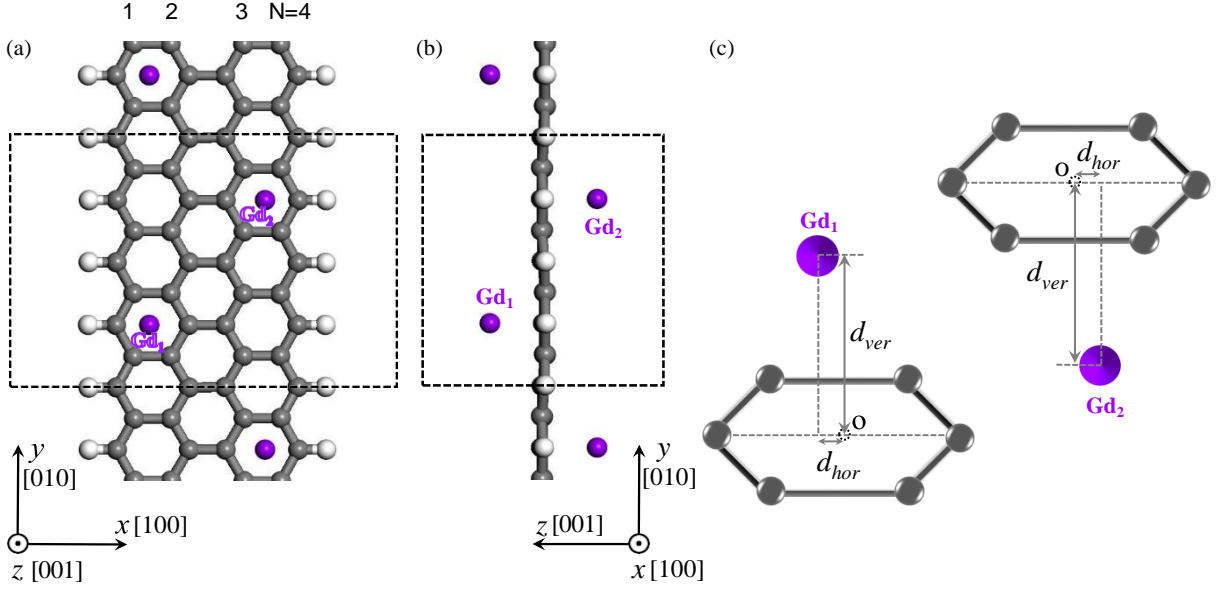

FIG. 1: (a) The top view (along $z-$axis [001]) and (b) the side view (along $x-$axis [100]) of the relaxed 2Gd-4ZGNR structure. (c) Schematic illustration of the parameters $d_{ver}$ and $d_{hor}$ that characterize the adsorption of the Gd adatoms horizontally deviating from the center of hexacarbon ring (point ′O′). The unit cell is indicated by broken lines in (a) and (b). Gray, purple and white balls represent carbon, gadolinium and hydrogen atoms, respectively. Note that there are only minor differences between the adsorption position in the FM and AFM states (see TABLE I).

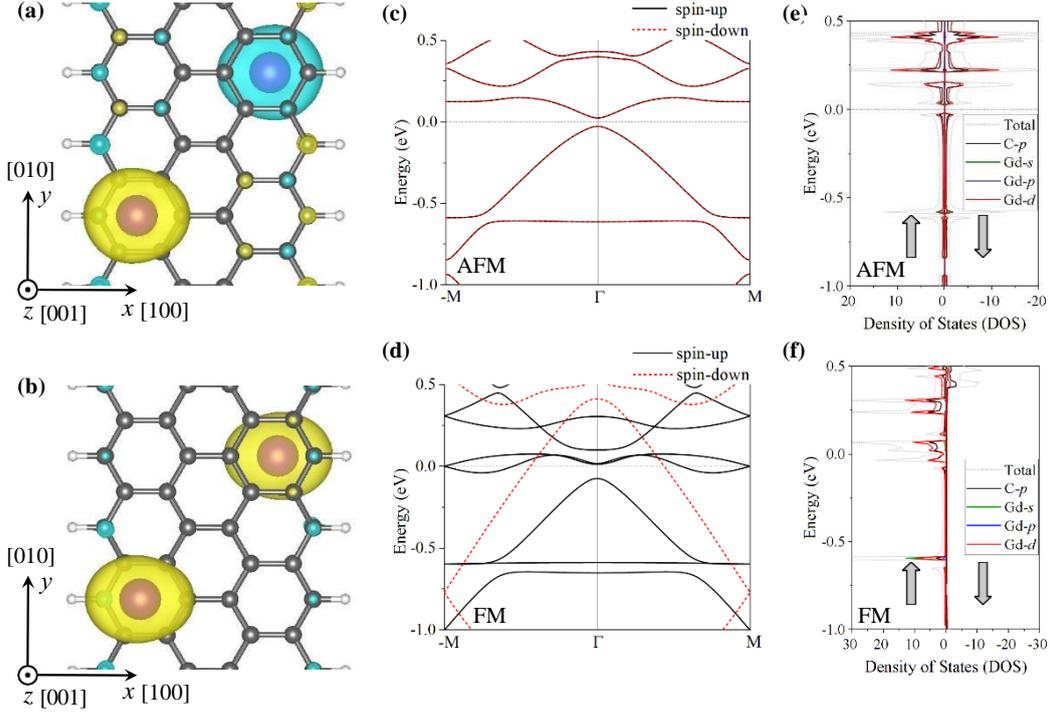

FIG. 2: Spin densities of the 2Gd-4ZGNR system for (a) the AFM state and (b) the FM state where the isosurface is plotted at $5\times10^{-3}$ *e/Bohr³* with the yellow and cyan colors representing the positive and negative value, respectively. The band structures without SOC perturbation for (c) the AFM state and (d) the FM state. The density of states (DOS) for (e) the AFM state and (f) FM state. The Fermi levels are set to zero in all the band structures and DOS.

The binding energy of the 2Gd-4ZGNR system with (A)FM state is defined as

$$E_b^{(A)FM} = 2E_{Gd} + E_{4ZGNR} - E_{2Gd-4ZGNR}^{(A)FM}, \qquad (3.1)$$

where the $E_{Gd}$ and $E_{4ZGNR}$ are the total energies of an isolated single Gd atom and the 4ZGNR composed of 32 carbon atoms and 8 hydrogen atoms, respectively, and the $E_{2Gd-4ZGNR}^{(A)FM}$ is the total energy of 2Gd-4ZGNR system in the (A)FM state. The binding energies and structural parameters that characterize the adsorption of the Gd adatoms are summarized in TABLE I for both the AFM and FM states. The parameters $d_{ver}$ and $d_{hor}$ are the vertical distance from a Gd adatom to the graphene plane and the horizontal offset to the center of the hexacarbon ring, respectively, shown in FIG. 1 (c). By comparing the adsorption parameters of the FM state to those of the AFM state, it is shown that the $d_{ver}$ slightly increases by 0.24% and the $d_{hor}$ remarkably increases by 57%. The large relative difference of the $d_{hor}$ reveals that the adsorption can be impacted by the magnetic states. As shown in FIG. 2 (a) and (b), in the AFM state, the two Gd adatoms show opposite spin directions and so do the two edges of the ZGNR. In addition, each Gd adatom is in a spin state opposite to that of the edge close to it for both the AFM and the FM states. The band structures in FIG. 2 (c) and (d) show that the AFM state is a semiconductor with a band gap of 50.52 meV and that the FM is a metal. The density of states (DOS) of the AFM and FM states in FIG. 2 (e) and (f) show that the Gd $d$-orbitals are coincident with the carbon $p$-orbitals in a broad energy interval, which implies a covalent bond between the Gd-adatom and the carbon atoms. The $4f$-shell of each Gd adatom is half-filled and contributes about 7 $\mu_B$ to the atomic magnetic moment. It is deeply buried in the valence band, as shown in FIG. S3 of Supplementary Information[41]. Hence, the Gd $d$−orbital plays the major role in determining the magnetic properties.

TABLE I: $E_b$ (eV), $d_{ver}$ (Å) and $d_{hor}$ (Å) of 2Gd-4ZGNR systems in the AFM and FM states.

|     | $E_b$(eV) | $d_{ver}$(Å) | $d_{hor}$(Å) |
|-----|-----------|--------------|--------------|
| AFM | 5.254     | 2.118        | 0.122        |
| FM  | 5.267     | 2.123        | 0.192        |

**B. Rashba effects of AFM and FM states**

As the SOC interaction is included with the spin quantization directions (SQD) along the $+z$ and $-z$ directions, respectively, as shown in FIG. 3 (a), or along the $+x$ and $-x$ directions, respectively, as shown in Fig. 3 (b), the Gd $d$−bands near the Fermi energy shift in the opposite directions along the $k$−axis on the band diagram of the AFM state, indicating a phenomenon associated with the Rashba effect[16, 17]. The Rashba Hamiltonian can be written as $H_R = \alpha_e(\sigma \times k) \cdot e$, where $\alpha_e$ is the Rashba parameter regulating the coupling strength between the wave vector ($k$) and the spin vector ($\sigma$) associated with the electric field direction ($e$). Both the $x-$ and $z-$components of the local built-in electric field are non-zero at the two Gd adatoms due to the off-center adsorption. By assuming that the dispersion is parabolic and that the spin quantization direction (SQD) is along the $+z$ direction, the resulting parabolic bands with the Rashba splitting due to the transverse electric field component ($E_x$) are given by $E(k_y) = \frac{\hbar^2}{2m^*}k_y^2 + \alpha_x k_y$, where $m^*$ and $k_y$ are the electric effective mass and the longitudinal wave number, respectively, and $\alpha_x$ is the Rashba parameter associated with the transverse electric field component $E_x$. This effect is coined transverse Rashba effect to emphasize that it is associated with the electric field component transverse to the ZGNR direction and to discriminate it from the effect associated with the electric field perpendicular to the ZGNR plane. The parabolic split by the transverse Rashba effect can be further rewritten as $E(k_y) = \frac{\hbar^2}{2m^*}(k_y + \Delta k_R)^2 - E_R$, where $\Delta k_R = \frac{m^* \alpha_x}{\hbar^2}$ and $E_R = \frac{m^* \alpha_x^2}{2\hbar^2}$ for electron dispersion are indicated in FIG. 3 (a). The Rashba parameter can be then extracted as $\alpha_x = 2E_R/\Delta k_R$. For the AFM state with the SQD along the $+z$ ([001]) direction, there is an obvious Rashba splitting of the valence band maximum (VBM) or conduction band minimum (CBM) at the $\Gamma$ point, from which it is extracted that the Rashba parameter $\alpha_x$ is 1.51 eVÅ. This value is comparable to those of the low-dimensional systems[46, 47], such as Pt-Si nanowire, with a giant Rashba parameter[47].

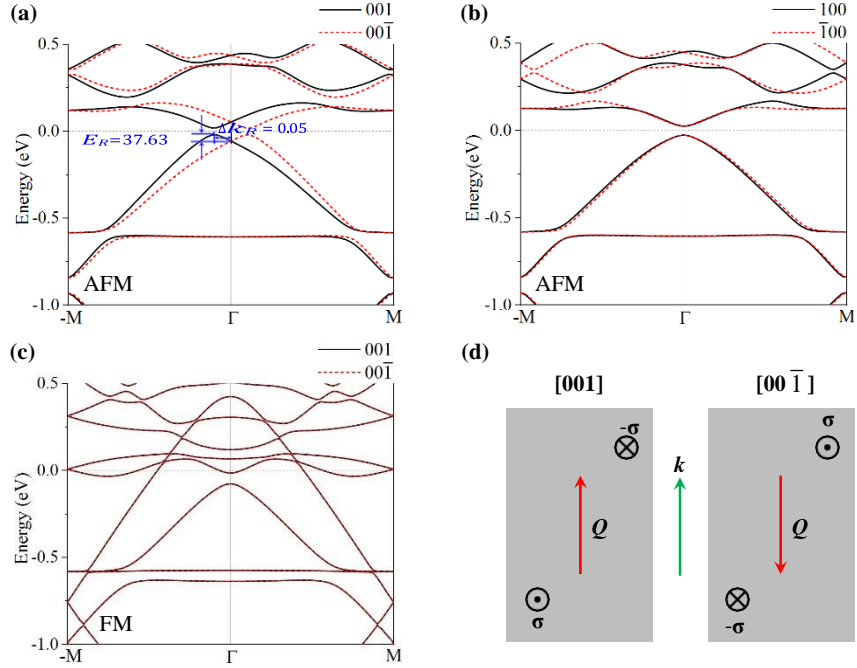

FIG. 3: The band structures with SOC-perturbation and magnetization along (a) [001] and [00$\bar{1}$] and (b) [100] and [$\bar{1}$00] directions for the AFM state, where the Rashba shifts in energy ($E_R$) and momentum ($\Delta k_R$) are indicated in (a). (c) Band structures with SOC perturbation and magnetization along [001] and [00$\bar{1}$] directions for the FM state. (d) Schematic illustrations of the directions of local spins, screw vector, and wavevector when spin quantization directions is along [001] and [00$\bar{1}$], respectively. The Fermi levels are set to zero in all the band structures.

When the SQD is along the $x$−axis, the Rashba splitting is associated with the $z$−component of the built-in electric field induced by the adsorption of the Gd adatoms, which apparently induces a much weaker splitting at the highest valence band and the lowest conduction band than that when the SQD is along the $z$−axis, as shown in FIG. 3 (b). Analyzing the reasons of this discrepancy, based on the $d$−orbital−resolved diagram of Gd atoms on band structure as shown in FIG. 6 (b), it is noticed that the zones contributed by $d_{xy}$, $d_{x^2-y^2}$ and $d_{xz}$ orbits of Gd atoms have an obvious relative large Rashba splitting for SQD along the $z$−axis, as shown in FIG. 3 (a). However, a much weaker Rashba splitting occurred in the zones where the bands dominated by $d_{z^2}$ and $d_{yz}$ orbits. On the contrary, for SQD along the $x$−axis corresponding to the $z$−component of the built-in electric field (FIG. 3 (b)), the zones mainly contributed by $d_{z^2}$, $d_{xz}$ and $d_{yz}$ orbits of Gd atoms have a large Rashba splitting, however, no apparent Rashba splitting observed in the zones where $d_{xy}$ and $d_{x^2-y^2}$ orbits have dominant contributions. Generally, there is no Rashba splitting in a certain band zone where the components of $d$ orbit of Gd atoms normal to the direction of electric field. Thus the conclusion can be obtained that the relations between the direction of the electric field and the distributions of the $d$-orbital components of Gd atoms on the band structure have a great influence on the magnitude of Rashba splitting at the corresponding band.

In contrast to the AFM state, there is no Rashba splitting in the FM state with the SQD along the $z$−axis (FIG. 3 (c)). The Rashba Hamiltonian of the 2Gd-4ZGNR system can be rewritten as

$$\boldsymbol{H}_{R2} = \alpha_{R,e}[(\boldsymbol{\sigma}_1 - \boldsymbol{\sigma}_2) \times \boldsymbol{k}] \cdot \boldsymbol{e} \; , \tag{3.2}$$

where the inversion symmetry ensures that the Rashba parameters are equal and the local electric field directions are opposite at the two Gd adatoms. For the AFM state, $\boldsymbol{\sigma}_1 = -\boldsymbol{\sigma}_2$, and thus the Rashba splitting at the two Gd adatoms enhance each other. For the FM state, however, $\sigma_1=\sigma_2$, and thus the Rashba splitting on the two Gd adatoms perfectly cancel each other out. As a result, there is no Rashba splitting for the FM state with the SQD along the $z$−axis. In fact, the Rashba splitting in the AFM state is consistent with the so-called R−2 type spin polarization proposed in the nonmagnetic 3D crystals with a centric symmetry, where local electric fields induce anti-parallel local spins on two sites associated with the inversion operation[8]. It should be noted that the 2Gd-4ZGNR systems in the AFM state also renders a centroic symmetry that was thought to be incompatible with the Rashba effect. The Rashba effect discovered in the AFM state can be described as the inverse effect of the R−2 type spin polarization in the nonmagnetic crystals, as it takes place for the anti-parallel spins of the AFM state.

In addition, it should be noted that the Rashba effect of the AFM state is distinct from the Rashba spin splitting of the nonmagnetic structures with or without a centric symmetry. In FIG. 3 (a) and (b), the up and down spins shift toward the same direction in the one-dimensional $k$−space, and the spin symmetry of the AFM state is fully preserved. This results from the centric symmetry of the 2Gd-4ZGNR system and is different from the Rashba spin splitting of the nonmagnetic structures, where the up and down spins are degenerate but separated by shifting oppositely in $k$-space. It is the SQD instead of the spin direction that is coupled to the crystalline momentum ($\hbar\mathbf{k}$) in the Rashba effect of the AFM state, as shown in FIG. 3 (a) and (b). It is more convention to rewrite (3.2) as

$$\boldsymbol{H}_{R2}=\alpha_{R,e}\boldsymbol{Q}\cdot\boldsymbol{k}, \tag{3.3}$$

where $\boldsymbol{Q}\equiv\boldsymbol{e}\times(\boldsymbol{\sigma}_1-\boldsymbol{\sigma}_2)$ is defined to be the screw vector of the 2Gd-4ZGNR. For the transverse Rashba effect of the AFM state when SQD is along the $+z$ ([001]) direction, the local spins on the two Gd adatoms are along the $z$−axis and antiparallel to each other, and the screw vector is along the $+y$ direction and parallel to the wavevector (FIG. 3 (d)). It follows that $\boldsymbol{H}_{R2}=2\alpha_{R,e}\sigma_z k_y$, which leads to a shift of both up and down spins toward the $-k$ direction in FIG. 3 (a). As the SQD flips from the $+z$ to the $-z$ direction, the local spins flip their directions, respectively, and the screw vector flips to the $-y$ direction (FIG. 3 (d)). It follows that $\boldsymbol{H}_{R2}=-2\alpha_{R,e}\sigma_z k_y$, which results in the shift toward the $+k$ direction in FIG 3 (a).

### C. The origin of magnetocrystalline anisotropy

The MCA for the AFM and FM states are calculated as a function of $\theta$ and fitted to a quadratic polynomial of $\cos\theta$, where $\theta$ is the polar angle of the magnetization in the $zy$−plane, as shown in FIG. 4. The uniaxial MCA constant ($K_u$) is calculated from the total energy difference between the in-plane [010] and out-of-plane [001] magnetization directions ($K_u = E_{[010]} - E_{[001]}$). The calculated $K_u$ of the AFM and FM states are 2.855 meV and 4.458 meV, respectively, and the out-of-plane (perpendicular) magnetization direction is preferred for both the states. We notice that the $K_u$ of the FM state is about twice that of the AFM state, which inspires us to explore the origin of the MCA of the two magnetic states.

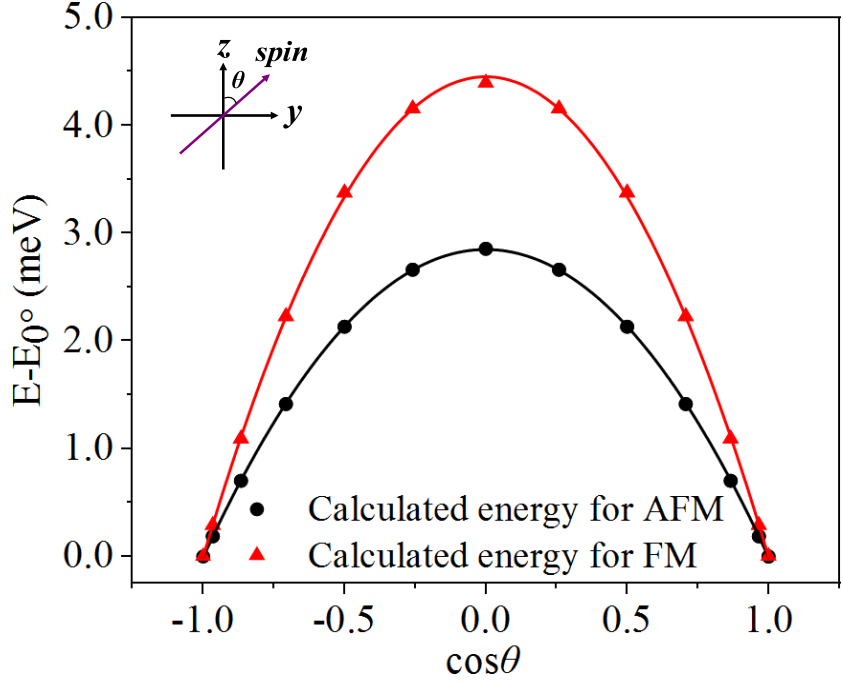

FIG. 4: Quadratic fitting of the calculated MCA (meV) versus $\cos\theta$ ($0° \leq \theta \leq 180°$) for the AFM (black line) and FM states (red line). The inset illustrates the definition of $\theta$. Fitting coefficients and standard errors are listed in TABLE. S I of Supplementary Information[41].

### 1. The AFM state

From the viewpoint of electronic structure, MCA results from the variation of band structure induced by rotating magnetization from one direction to another. It can thus be written as a summation in $k$−space[48, 49],

$$\text{MCA} = E(a) - E(b) = \sum_{k \in FBZ} \{ \sum_{i \in NBANDS} [\varepsilon_i^a(k) f_i^a(k) - \varepsilon_i^b(k) f_i^b(k)] * w(k) \} \qquad (3.4)$$

where $E(a)$ and $E(b)$ represents the energy when the magnetization along $a$ and $b$ directions,

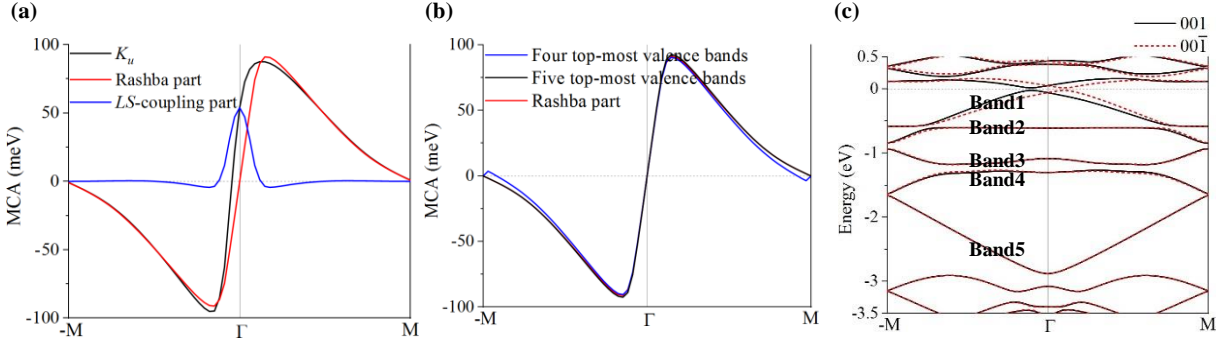

FIG. 5: For the AFM state. (a) The distributions of the $K_u$, the Rashba part, and the $LS$−coupling part in $k$−space. (b) The Rashba part of $K_u$ distribution in $k$−space compared to that calculated in terms of (3.4) with the four or five top-most valence bands. (c) The labels of the five top-most valence bands.

respectively. $\varepsilon(k)$, $f(k)$, and $w(k)$ are the band energy, the filling factor, and the weight at a point $k$ in the FBZ, and the subscript $i$ stands for the $i$th band and the superscripts $a$ and $b$ denote the magnetization directions. The $K_u$ of the AFM state is calculated to be 2.857 meV in terms of (3.4) with $a$ and $b$ set to be [010] and [001] directions respectively and $i$ contains all the bands, which is almost equal to that calculated directly from the total energy difference. In addition, the contribution to the $K_u$ at a single $k$−point can be extracted from (3.4) and plotted as a function of $k$, as shown by the black curve in FIG. 5 (a), which is approximately antisymmetric about the Γ point like the result of the 4ZGNR decorated with a single Gd adatom[22]. In fact, the $k$−space distribution of $K_u$ can be decomposed into the odd part associated to Rashba splitting and the even part associated with $LS$−coupling for the AFM state. By setting $a$ and $b$ to be [00$\bar{1}$] and [001] directions respectively, in (3.4) the $LS$−coupling part of $K_u$ vanishes, the Rashba splitting part is calculated and plotted, as the red curve in FIG. 5 (a), which is rigorously an odd function of $k$. As a result, the net contribution to the $K_u$ from the Rashba splitting vanishes. In fact, the transverse Rashba effect of the AFM state causes the opposite shifts of the bands about the Γ−point when the magnetization directions are set to be [00$\bar{1}$] and [001], resulting in that the Rashba part of $K_u$ is a rigorous odd function of $k$ in the one-dimensional FBZ. By comparing the Rashba part of $K_u$ distribution in $k$−space calculated with all the bands of the AFM state to those calculated with the four or five top-most valence bands (FIG. 5 (c)). As shown in FIG. 5 (b), it is shown that the five top-most valence bands approximately reproduce the Rashba part.

After the $LS$−coupling part is isolated from the $K_u$ in $k$−space as shown by the blue curve in FIG. 5 (a), the origin of MCA can be analyzed in terms of perturbation theory[48]. Both the first and second order perturbations can lower down the energy of filled states, depending on magnetization direction, and result in MCA. The first order perturbation of $LS$-coupling acts on the degenerate or nearly degenerate states at the Fermi energy, and open or enlarge the gap, from which the $LS$-coupling coefficient can be extracted. Although the second order perturbation can also contribute to the MCA in general, it is less important for one-dimensional systems, since its contribution is a half order higher than that of the first order, as shown in the section Introduction.

The $LS$-coupling part of $K_u$ as a function of $k$ without the weight $w(k)$ in (3.4) for the AFM state is plotted in FIG. 6 (a). To isolate the $LS$-coupling effect in the band structure, a set of bands is defined by averaging the band structures calculated with the magnetization along the [001] and [00$\bar{1}$] directions respectively, that is, each band of this set is defined by $\varepsilon_i(k) = (\varepsilon_i^{[001]}(k) + \varepsilon_i^{[00\bar{1}]}(k))/2$. It should be noted that there is no Rashba effect in the averaged band structure, as shown by the bands in black in FIG. 6 (c), because the Rashba splitting associated with the [001] and [00$\bar{1}$] magnetization directions are opposite to each other and canceled by the averaging. In addition, the band structure calculated with the magnetization direction along the [010] is plotted in FIG. 6 (c) as a comparison. In FIG. 6 (d), it is obvious that the bandgap is enlarged from 49.07 meV to 109.20 meV by rotating the magnetization direction from the longitudinal to the perpendicular direction with respect to the ZGNR. This first-order perturbation of $LS$-coupling interaction lowers down the VBM of the averaged band structure, compared to the [010] band structure, and results in the sharp peak near the $\Gamma$ point (or along the $ab$ segment) in the $LS$-coupling part of $K_u$ (FIG. 6 (a)). In fact, the $LS$-coupling term in Hamiltonian can be written as

$$\mathbf{H}_{ls} = \lambda \mathbf{l} \cdot \mathbf{s} = \lambda(l_z \cos\theta + l_y \sin\theta)s = \frac{\lambda}{2}(l_z \cos\theta + l_y \sin\theta), \tag{3.5}$$

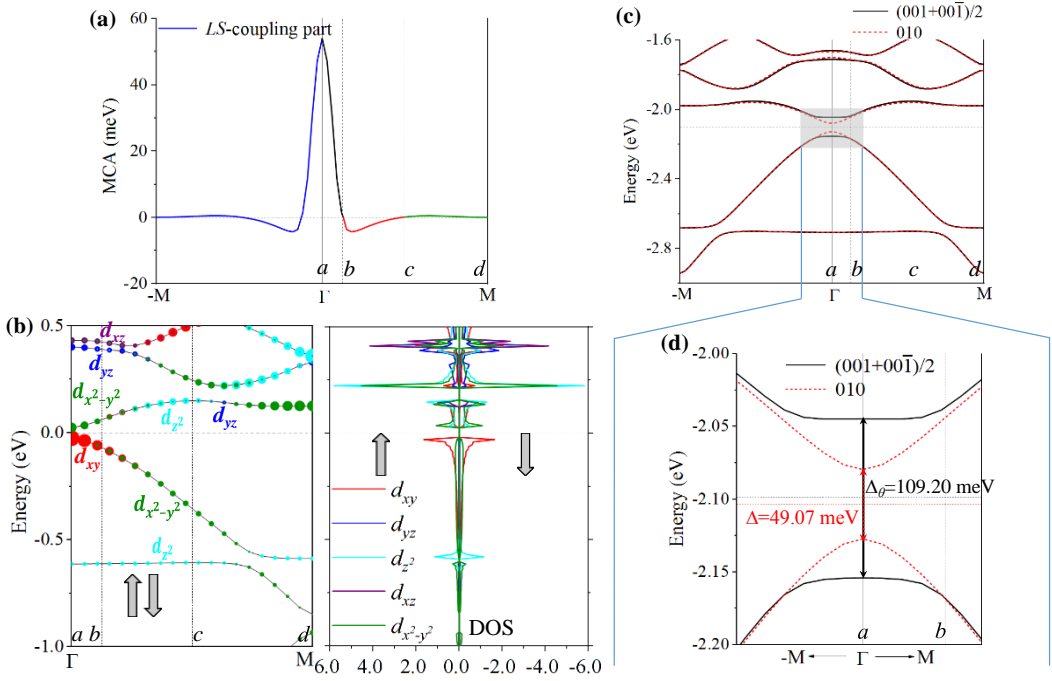

FIG. 6: All subplots are associated with the AFM state of 2Gd-4ZGNR system. (a) $LS$−coupling part of $K_u$ distribution without considering weight factor $w(k)$ in (3.4) in $k$−space which is divided into three segments (labeled as $ab$, $bc$ and $cd$) along the Γ−M line. (b) Spin-polarized band structure (left) and projected DOS of Gd $5d$−orbitals where the markers $a$, $b$, $c$, $d$ along the Γ−M line are those in (a). (c) Band structures with SOC perturbation, where the bands in red are calculated with the magnetization along the [010] direction and those in black are the average of the bands calculated with the magnetization along the [001] and the [00$\bar{1}$] directions, respectively, that is, for the $i$th band at point $k$, $\varepsilon_i(k) = (\varepsilon_i^{[001]}(k) + \varepsilon_i^{[00\bar{1}]}(k))/2$. The markers $a$, $b$, $c$, $d$ along the Γ−M line are those in (a). (d) Zoom-in view of the gray area of (c), where the corresponding band gaps are marked.

where the electron spin ($s = 1/2$) is in the $zy$−plane, $\lambda$ is the $LS$ coupling coefficient, and $\theta$ is the polar angle of the spin. The band structure projected onto the $d$−orbitals is plotted in FIG. 6 (b), where the VBM and CBM at the Γ point are essentially contributed by $d_{xy}$ and $d_{x^2−y^2}$ orbits, respectively. The matrix representation of (3.5) can then be constructed by using $d_{xy}$ and $d_{x^2−y^2}$ orbits as

$$[H_{ls}] = \begin{bmatrix} \varepsilon & i\lambda\cos\theta \\ -i\lambda\cos\theta & \varepsilon + \Delta \end{bmatrix}, \quad (3.6)$$

where $\varepsilon$ and $\varepsilon + \Delta$ are the VBM and CBM energy levels without the $LS$−coupling per-turbation at the Γ point respectively, and Δ is the bandgap without the perturbation. By diagonalizing the matrix in (3.6), two levels with the perturbation can be obtained and separated by an enlarged $\theta$-dependent gap

$$\Delta_\theta = \sqrt{\Delta^2 + 4(\lambda \cos \theta)^2}, \tag{3.7}$$

By using the gaps calculated by the first-principles, Δ=49.07 meV and $\Delta_\theta$=109.20 meV at $\theta$=0°, the LS coupling coefficient $\lambda$ can be calculated for the AFM state of the 2Gd-4ZGNR system to be 48.78 meV, which is close to the value of about 60 meV in Gd-adsorbed graphene[45]. In fact, there is no Rashba splitting in the band structure calculated with SOC interaction when the magnetization is along the [010] direction, as the spin direction is parallel to the wave vector. In addition, there is also no first-order LS−coupling perturbation at the Γ point, as the off-diagonal elements in (3.7) at $\theta$=90° vanish (TABLE. S II of Supplementary Information[41]).

The contribution to the $K_u$ from the second order perturbation of LS−coupling interaction[48] is further discussed qualitatively along the bc and cd segments in FIG. 6 (a). It is given by the difference between the second order perturbation energies with the magnetization along the [010] and [001] directions respectively, that is

$$K_u^{(2)}(k) = \Delta E^{(2)}(k;[010]) - \Delta E^{(2)}(k;[001]) = \xi^2 \sum_{o,u} \frac{|\langle o|L_z|u\rangle|^2 - |\langle o|L_y|u\rangle|^2}{\varepsilon_u - \varepsilon_o}, \tag{3.8}$$

where $\xi$ is the LS−coupling coefficient that is determined from the bandgap enlarged by the first-order perturbation at the Γ point, and the summation is over all occupied (o) and unoccupied (u) states, both of which are either spin-up or spin-down. The matrix elements of angular momentum operators ($L_x$, $L_y$, and $L_z$) are calculated with the cubic harmonics of d−orbitals and summarized in TABLE. S II of Supplementary Information[41].

Several approximations can be adopted to simplify the analysis of the second-order perturbation in terms of (3.8) where all transitions are considered between the occupied and unoccupied states. In fact, the MCA of an insulating magnet is associated with the localized $d$−orbitals of transition metal ions, and thus only the transitions involving the $d$−orbitals of a single Gd adatom need to be considered. In addition, the exchange splitting (the separation in energy) between the spin-up and spin-down $d$−orbitals of a single Gd adatom is as large as 1.86 eV in the AFM state (shown in FIG. S4 (a) of Supplementary Information[41]), which goes into the denominator of equation (3.8) and is two orders of magnitude stronger than the $LS$−coupling coefficient $\xi$ in the numerator. Thus, the transitions between the states in opposite spins can be neglected. More than that, the occupied $d$−orbitals of the two Gd adatoms are associated with the spin-up and spin-down channels, respectively, in the AFM state (shown in FIG. S4 (b) of Supplementary Information[41]). Thus, it is only necessary to consider either spin-up or spin-down $d$−bands near the bandgap.

The contribution of the second-order perturbation to the $K_u$ in $k$−space (FIG. 6 (a)) is discussed in terms of (3.8) with the transitions between the spin-up (or spin-down) $d$−orbitals near the bandgap (FIG. 6 (b)). Along the segment $ab$, the major contribution is the transition over the narrow bandgap near the Γ point between the highest valence band and the lowest conduction band that are mainly of $d_{xy}$ and $d_{x^2-y^2}$ characters, respectively. This contribution is positive and partly canceled by the negative contribution associated with the transition between the $d_{xy}$−component of the highest valence band and the $d_{yz}$−component of the second lowest conduction band. Along the $bc$ segment, there are positive and negative contributions competing with each other, which results in a weak negative $K_u$. The positive contribution is associated with the transition between the highest valence band and the lowest conduction band near the $b$ point that are both a mixture of $d_{xy}$ and $d_{x^2-y^2}$ characters. The negative contribution is associated with the transition between the $d_{x^2-y^2}$−component of the highest valence band and the $d_{xz}$−component of the third lowest conduction band. It also comes from the transition between the $d_{xy}$−component of the highest valence band and the $d_{yz}$−component of the second lowest conduction band near the $b$ point. The magnitude of $K_u$ is diminishing toward the $c$ point, as the denominator ($E_u-E_o$) in (3.8) increases due to the decrease of the highest valence band. Along the $cd$ segment, the $K_u$ is very weak due to the large $E_u-E_o$ in the denominator of (3.8).

**2. The FM state**

In FIG. 7 (a), the $LS$−coupling part of $K_u$ without the weight $w(k)$ in (3.4) is illustrated in $k$−space for the FM state. The MCA calculated in terms of (3.4) that numerically integrates the $k$−space distribution is 4.393 meV, almost equal to that determined from the difference between the total energies with the magnetization along [010] and [001] directions respectively (4.394 meV).

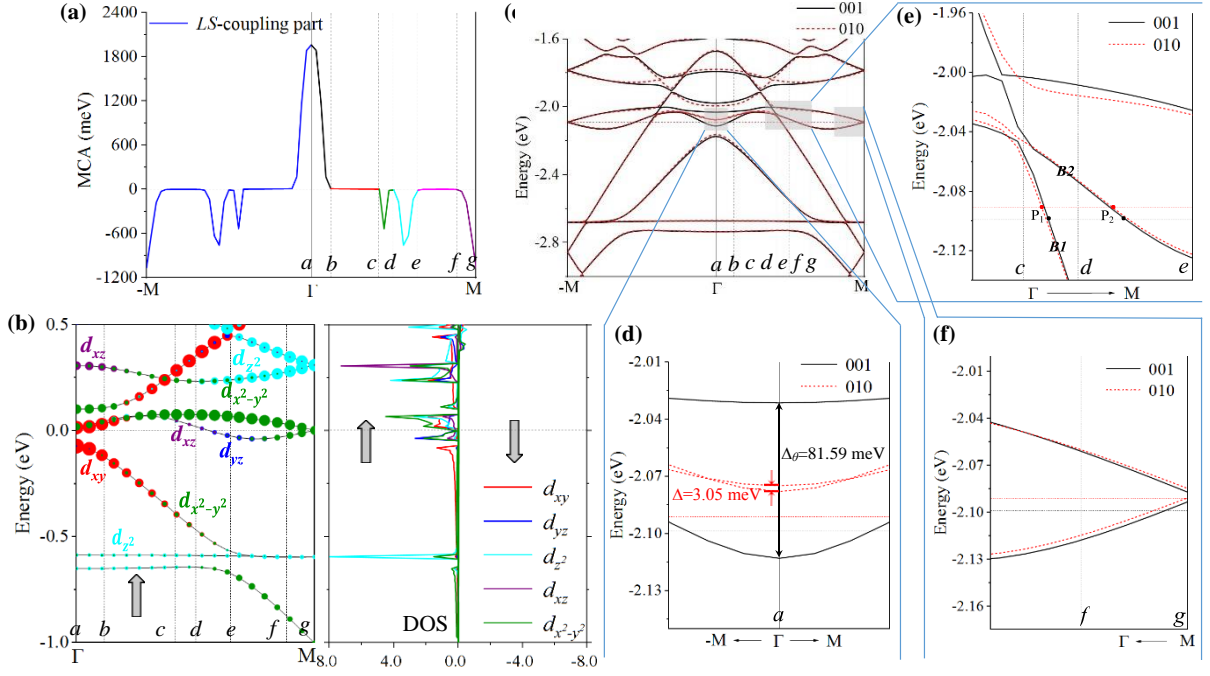

FIG. 7: All subplots are associated with the FM state of 2Gd-4ZGNR system. (a) $LS-$coupling part of $K_u$ as a function of $k$ without considering weight factor $w(k)$ in (3.4), where the spikes are marked out by the labels $a-g$ along the $k-$axis. (b) Spin-up band structure with the weights of Gd $5d-$orbital states near the Fermi energy (left), and associated projected DOS (right). The markers $a-g$ along the $\Gamma-M$ line are those in (a). (c) Band structures calculated with SOC perturbations for the [001] and [010] magnetization directions, respectively. The markers $a-g$ along the $\Gamma-M$ line are those in (a). (d), (e) and (f) correspond to the zoom-in view of the gray area of (c), respectively, where the corresponding band gaps in (d) are marked, the labels $P_1$ ($P_2$) in (e) are the crossover points between the band B1 (B2) of the [001] or [010] magnetization directions and the corresponding Fermi energy.

There are several spikes, either positive or negative, in FIG. 7 (a), which results from the drastic change of the filling factor in (3.4) at the spikes due to the shift of Fermi energy induced by rotating magnetization direction. This effect is well-known for calculating the MCA of metals in terms of the force theorem[29, 30], and might lead to the numerical instability of the result for a cubic metal, of which is the MCA usually three order of magnitude lower than that calculated here. The so-called state-tracking method[31] was proposed to solve this problem, and trigged a debate over the validity of the method[34, 50]. In fact, the shift of Fermi energy is induced by the first-order perturbation of the $LS$-coupling interaction between the almost-degenerate lowest and second lowest conduction bands at the $\Gamma$ point, when the magnetization is along the [001] direction (FIG. 7 (c)). Those two bands are mainly projected onto the $d_{xy}$ and $d_{x^2-y^2}$ orbits, and the first-order perturbation enlarges the pseudo-gap in terms of (3.7). From the pseudo-gap the $LS$-coupling coefficient ($\lambda$) is estimated to be 40.77 meV by substituting the *ab-initio* values $\Delta = 3.05$ meV and $\Delta_\theta = 81.59$ meV at $\theta = 90$ into (3.7) (shown in FIG. 7 (d)). Note that the off-diagonal elements of (3.6) vanish when the magnetization along [010] direction. The enhanced pseudo-gap pushes the local minimum of the lowest conduction band at the $\Gamma$ point downward to the energy lower than the Fermi energy. This effect converts the lowest unoccupied states around the $\Gamma$ point to the occupied ones when the magnetization is rotated from the [010] direction to the [001] direction, and thus induces the positive spike at the $\Gamma$ point.

The lowering-down of the lowest conduction band around the $\Gamma$ point not only directly induces the positive spike at the $\Gamma$ but also induces the negative spikes indirectly by shifting the Fermi energy. The parts of the band structure (grey-background boxes in FIG. 7 (c)) that are associated with the spikes in FIG. 7 (a) are zoomed in (FIG. 7 (d), (e), and (f)). In FIG. 7 (f), the Fermi energy notably shifts down and the bands near the Fermi energy slightly nudge, as the magnetization is rotated from [010] to [001] direction. This effect converts the occupied states near the M for the [010] magnetization direction to the unoccupied states for the [001] direction, and thus induces the negative spike along the *fg* segment in FIG. 7 (a) according to (3.4). This effect converts the occupied states near the $P_1$ and $P_2$ points (FIG. 7 (e)) for the [010] magnetization direction to the unoccupied states for the [001] direction, and thus induces the negative spikes along the *cd* and *de* segments in FIG. 7 (a).

## IV. CONCLUSION

The zigzag graphene nanoribbon decorated with two Gd adatoms per unit cell has been investigated by using the first-principle calculations based on the density functional theory. The 2Gd-4ZGNR system feature of the anti-plane adsorption structure and the inversion symmetry is energetically stable. Its magnetic ground state is AFM, which is lower than the metastable FM state by 45.383 meV. The off-center adsorption of the Gd adatoms generates the local dipole fields in the transverse direction with respect to the nanoribbon, and consequently induces the transverse Rashba effect that has been identified from the splitting of the SOC-included band structure for the AFM state. It is shown that the transverse Rashba effect at the two Gd adatoms enhances each other due to the inversion symmetry in the AFM state, and cancels out each other in the FM state. The Rashba splitting in the AFM state can be understood as the inverse effect of so-called R−2 type spin polarization that is AFM and generated by the Rashba effect in some centrosymmetric bulk crystals. It should be noted that both up and down spins shift toward the same direction in the SOC band structure and that the direction is associated with the screw vector of the quasi-one-dimensional AFM state. The transverse Rashba parameter ($\alpha_x$) 1.51 eVÅ from the splitting in the SOC band structure, comparable with the values of other one-dimensional Rashba systems.

The origin of MCA has been explored for the AFM and FM states, both of which shows a perpendicular MCA in an order of magnitude of 1 meV per Gd adatom. The uniaxial MCA constant ($K_u$) is determined by the quadratic polynomial fitting. The $k$−space distribution of $K_u$ is calculated for the [010], [001], and [00$\bar{1}$] magnetization directions, from which the Rashba and $LS$−coupling contributions to $K_u$ is separated in $k$−space. The Rashba part is an odd function of $k$ and its net contribution to total $K_u$ vanishes. The $LS$−coupling part is however an even function of $k$ and does contribute to the total $K_u$. It has been analyzed in terms of the SOC band structures. For the AFM state, the positive spike at the Γ point is attributed to the first-order perturbation between the highest valence band and the lowest conduction band that pushes the former down in energy. The second-order perturbation contribution is qualitatively discussed in terms of the band structure projected onto $d$−orbitals. For the FM state, the band structure is metallic. The first-order perturbation opens a pseudo-gap between the lowest conduction band and the second lowest one, which pushes the former down below the Fermi energy. This effect induces a puddle to the valence band at the Γ, and consequently the positive spike to the $LS$−coupling contribution in $k$−space. More than that, the puddle causes a shifting-down of Fermi energy and converts the occupied states near the Fermi energy to the unoccupied ones, which induces several negative spikes to the $LS$−coupling contribution in $k$−space. This work introduces a new type of Rashba effect in a quasi-one-dimensional antiferromagnet based on graphene nanoribbon, which couples the perpendicular magnetic easy direction to the wave-vector.

This finding implies a new approach to realize and control high coherency spin transportation in one dimensionality using graphene-based spintronics.

**V. ACKNOWLEDGMENTS**

This research is supported by the science Challenge Project (Grant No. TZ2016003-1- 105); Tianjin Natural Science Foundation (Grant No. 20JCZDJC00750); CAEP Microsystem and THz Science and Technology Foundation (Grant No. CAEPMT201501); National Basic Research Program of China (Grant No. 2011CB606405).

[1] Wolf, S. A., Awschalom, D. D., Buhrman, R. A., Daughton, J. M., von Moln*a*r, V. S., Roukes, M. L., ... and Treger, D. M. (2001). Spintronics: a spin-based electronics vision for the future. Science, 294(5546), 1488-1495.

[2] Bychkov, Y. A., and Rashba, *É*. I. (1984). Oscillatory effects and the magnetic susceptibility of carriers in inversion layers. Journal of physics C: Solid State Physics, 17(33), 6039.

[3] Bychkov, Y. A., and Rashba, *É*. I. (1984). Properties of a 2D electron gas with lifted spectral degeneracy. JETP Lett, 39(2), 78.

[4] Winkler, R. (2003). Spin-orbit coupling effects in two-dimensional electron and hole systems. Springer Tracts in Modern Physics, 191, 1-8.

[5] Ishizaka, K., Bahramy, M. S., Murakawa, H., Sakano, M., Shimojima, T., Sonobe, T., ... and Miyamoto, K. (2011). Giant Rashba-type spin splitting in bulk BiTeI. Nature Materials, 10(7), 521-526.

[6] Sakano, M., Miyawaki, J., Chainani, A., Takata, Y., Sonobe, T., Shimojima, T., ... and Nagaosa, N. (2012). Three-dimensional bulk band dispersion in polar BiTeI with giant Rashba-type spin splitting. Physical Review B, 86(8), 085204.

[7] Ast, C. R., Henk, J., Ernst, A., Moreschini, L., Falub, M. C., Pacil, D., ... and Grioni, M. (2007). Giant spin splitting through surface alloying. Physical Review Letters, 98(18), 186807.

[8] Zhang, X., Liu, Q., Luo, J. W., Freeman, A. J., and Zunger, A. (2014). Hidden spin polarization in inversion-symmetric bulk crystals. Nature Physics, 10(5), 387-393.

[9] Riley, J. M., Mazzola, F., Dendzik, M., Michiardi, M., Takayama, T., Bawden, L., ... and

Kim, T. K. (2014). Direct observation of spin-polarized bulk bands in an inversion-symmetric semiconductor. Nature Physics, 10(11), 835-839.

[10] Liu, Q., Zhang, X., Jin, H., Lam, K., Im, J., Freeman, A. J., and Zunger, A. (2015). Search and design of nonmagnetic centrosymmetric layered crystals with large local spin polarization. Physical Review B, 91(23), 235204.

[11] S-lawi$n$ska, J., Narayan, A., and Picozzi, S. (2016). Hidden spin polarization in nonmagnetic centrosymmetric $BaNiS_2$ crystal: Signatures from first principles. Physical Review B, 94(24), 241114.

[12] Wu, S. L., Sumida, K., Miyamoto, K., Taguchi, K., Yoshikawa, T., Kimura, A., ... and Tanaka, I. (2017). Direct evidence of hidden local spin polarization in a centrosymmetric superconductor $LaO_{0.55}F_{0.45}BiS_2$. Nature Communications, 8(1), 1-7.

[13] Gotlieb, K., Lin, C. Y., Serbyn, M., Zhang, W., Smallwood, C. L., Jozwiak, C., ... and Lanzara, A. (2018). Revealing hidden spin-momentum locking in a high-temperature cuprate superconductor. Science, 362(6420), 1271-1275.

[14] Yuan, L., Liu, Q., Zhang, X., Luo, J. W., Li, S. S., and Zunger, A. (2019). Uncovering and tailoring hidden Rashba spin−orbit splitting in centrosymmetric crystals. Nature Communications, 10(1), 1-8.

[15] Tu, J., Chen, X. B., Ruan, X. Z., Zhao, Y. F., Xu, H. F., Chen, Z. D., ... and Zhang, Y. (2020). Direct observation of hidden spin polarization in $2H-MoTe_2$. Physical Review B, 101(3), 035102.

[16] Krupin, O., Bihlmayer, G., Starke, K., Gorovikov, S., Prieto, J. E., Dürich, K., ... and Kaindl, G. (2005). Rashba effect at magnetic metal surfaces. Physical Review B, 71(20), 201403.

[17] Krupin, O., Bihlmayer, G., Dürich, K. M., Prieto, J. E., Starke, K., Gorovikov, S., ... and Kaindl, G. (2009). Rashba effect at the surfaces of rare-earth metals and their monoxides. New Journal of Physics, 11(1), 013035.

[18] Dedkov, Y. S., Fonin, M., Rüdiger, U., and Laubschat, C. (2008). Rashba effect in the graphene/Ni (111) system. Physical Review Letters, 100(10), 107602.

[19] Park, J. H., Kim, C. H., Lee, H. W., and Han, J. H. (2013). Orbital chirality and Rashba interaction in magnetic bands. Physical Review B, 87(4), 041301.

[20] Kawano, M., Onose, Y., and Hotta, C. (2019). Designing Rashba−Dresselhaus effect in magnetic insulators. Communications Physics, 2(1), 1-8.

[21] Yamauchi, K., Barone, P., and Picozzi, S. (2019). Bulk Rashba effect in multiferroics: A theoretical prediction for $BiCoO_3$. Physical Review B, 100(24), 245115.

[22] Qin, Z., Qin, G., Shao, B., and Zuo, X. (2017). Unconventional magnetic anisotropy in one-dimensional Rashba system realized by adsorbing Gd atom on zigzag graphene nanoribbons. Nanoscale, 9(32), 11657-11666.

[23] Qin, Z., Qin, G., Shao, B., and Zuo, X. (2020). Rashba spin splitting and perpendicular magnetic anisotropy of Gd-adsorbed zigzag graphene nanoribbon modulated by edge states under external electric fields. Physical Review B, 101(1), 014451.

[24] Shick, A. B., Khmelevskyi, S., Mryasov, O. N., Wunderlich, J., and Jungwirth, T. (2010). Spin-orbit coupling induced anisotropy effects in bimetallic antiferromagnets: A route towards antiferromagnetic spintronics. Physical Review B, 81(21), 212409.

[25] Jungwirth, T., Marti, X., Wadley, P., and Wunderlich, J. (2016). Antiferromagnetic spintronics. Nature Nanotechnology, 11(3), 231-241.

[26] Baltz, V., Manchon, A., Tsoi, M., Moriyama, T., Ono, T., and Tserkovnyak, Y. (2018). Antiferromagnetic spintronics. Reviews of Modern Physics, 90(1), 015005.

[27] Van Vleck, J. H. (1937). On the anisotropy of cubic ferromagnetic crystals. Physical Review, 52(11), 1178.

[28] Lessard, A., Moos, T. H., and Hübner, W. (1997). Magnetocrystalline anisotropy energy of transition-metal thin films: A nonperturbative theory. Physical Review B, 56(5), 2594.

[29] Weinert, M., Watson, R. E., and Davenport, J. W. (1985). Total-energy differences and eigenvalue sums. Physical Review B, 32(4), 2115.

[30] Daalderop, G. H. O., Kelly, P. J., and Schuurmans, M. F. H. (1990). First-principles calculation of the magnetocrystalline anisotropy energy of iron, cobalt, and nickel. Physical Review B, 41(17), 11919.

[31] Wang, D. S., Wu, R., and Freeman, A. J. (1993). State-tracking first-principles determination of magnetocrystalline anisotropy. Physical Review Letters, 70(6), 869.

[32] Wang, X., Wu, R., Wang, D. S., and Freeman, A. J. (1996). Torque method for the theoretical determination of magnetocrystalline anisotropy. Physical Review B, 54(1), 61.

[33] Schneider, G., Erickson, R. P., and Jansen, H. J. F. (1997). Calculation of the magnetocrystalline anisotropy energy using a torque method. Journal of applied physics, 81(8), 3869-3871.

[34] Bylander, D. M., and Kleinman, L. (1995). Effect of $k$-space integration on self-consistent

surface magnetic anisotropy calculations. Physical Review B, 52(3), 1437.

[35] Yang, I., Savrasov, S. Y., and Kotliar, G. (2001). Importance of correlation effects on magnetic anisotropy in Fe and Ni. Physical Review Letters, 87(21), 216405.

[36] Blöchl, P. E. (1994). Projector augmented-wave method. Physical Review B, 50(24), 17953.

[37] Kresse, G., and Furthmüller, J. (1996). Efficiency of ab-initio total energy calculations for metals and semiconductors using a plane-wave basis set. Computational Materials Science, 6(1), 15-50.

[38] Perdew, J. P., Burke, K., and Ernzerhof, M. (1996). Generalized gradient approximation made simple. Physical Review Letters, 77(18), 3865.

[39] Dudarev, S. L., Botton, G. A., Savrasov, S. Y., Humphreys, C. J., and Sutton, A. P. (1998). Electron-energy-loss spectra and the structural stability of nickel oxide: An LSDA+U study. Physical Review B, 57(3), 1505.

[40] Wang, B., Zhang, X., Zhang, Y., Yuan, S., Guo, Y., Dong, S., and Wang, J. (2020). Prediction of a two-dimensional high-TC $f$-electron ferromagnetic semiconductor. Materials Horizons, 7, 1623-1630

[41] Supplementary Information is available online

[42] Makov, G., and Payne, M. C. (1995). Periodic boundary conditions in *ab* initio calculations. Physical Review B, 51(7), 4014.

[43] Son, Y. W., Cohen, M. L., and Louie, S. G. (2006). Energy gaps in graphene nanoribbons. Physical Review Letters, 97(21), 216803.

[44] Wu, F., Kan, E., Xiang, H., Wei, S. H., Whangbo, M. H., and Yang, J. (2009). Magnetic states of zigzag graphene nanoribbons from first principles. Applied Physics Letters, 94(22), 223105.

[45] Lu, Y., Zhou, T. G., Shao, B., Zuo, X., and Feng, M. (2016). Carrier-dependent magnetic anisotropy of Gd-adsorbed graphene. AIP Advances, 6(5), 055708.

[46] Barke, I., Zheng, F., Rügheimer, T. K., and Himpsel, F. J. (2006). Experimental evidence for spin-split bands in a one-dimensional chain structure. Physical Review Letters, 97(22), 226405.

[47] Park, J., Jung, S. W., Jung, M. C., Yamane, H., Kosugi, N., and Yeom, H. W. (2013). Self-assembled nanowires with giant rashba split bands. Physical Review Letters, 110(3), 036801.

[48] Wang, D. S., Wu, R., and Freeman, A. J. (1993). First-principles theory of surface magnetocrystalline anisotropy and the diatomic-pair model. Physical Review B, 47(22), 14932.

[49] Van der Laan, G. (1998). Microscopic origin of magnetocrystalline anisotropy in transition metal thin films. Journal of Physics: Condensed Matter, 10(14), 3239.

[50] Daalderop, G. H. O., Kelly, P. J., and Schuurmans, M. F. H. (1993). Comment on State-tracking first-principles determination of magnetocrystalline anistropy. Physical Review Letters, 71(13), 2165.

## Supplementary Information

It is established that narrow zigzag graphene nanoribbon (ZGNR) is antiferromagnetic (AFM) semiconductors, where each edge is a ferromagnetic (FM) chain and the two edges are antiferrmagnetically coupled each other[1, 2]. The magnetic states of the ZGNR composed of four zigzag carbon chains (4ZGNR) are investigated to verify the computational approach. The edge carbon atoms are saturated by hydrogens. It is shown that the AFM state is the magnetic ground state, which is lower than the FM state by 14.588 meV per edge carbon atom in energy and lower than the non-magnetic state by 55.436 meV, which agrees well with the previous calculations[3, 4].

In Fig. S1, the number of $k$-points can explicitly influence on the MCA. We find $1 \times 69 \times 1$ $k$-points along 1D $k$-space is sufficient to calculate the accurate MCA of 2Gd-4ZGNR system for AFM state.

In Fig. S2, the favorable configuration of 2Gd-4ZGNR system corresponding to a coverage of two Gd adatoms per 16 carbon atoms and the band structures of non-SOC perturbation as well as [001] and [00$\bar{1}$] magnetization directions are shown in Fig. S2. The Rashba-type splitting is shown when the magnetization directions along [001] and [00$\bar{1}$] axis for AFM state. The calculational binding energy, Rashba parameter along [001] ([00$\bar{1}$]) magnetization direction and MCA are 5.085 eV, 0.958 eV Å and 2.240 meV, respectively.

In Fig. S3, we can clearly see that the distributions of 4$f$-orbital states of the Gd atoms in band structure are far away from the fermi level, which means that 4$f$-orbital states have little influence on the properties of whole system.

[1] Son, Y. W., Cohen, M. L., and Louie, S. G. (2006). Energy gaps in graphene nanoribbons. Physical Review Letters, 97(21), 216803.

[2] Wu, F., Kan, E., Xiang, H., Wei, S. H., Whangbo, M. H., and Yang, J. (2009). Magnetic states of zigzag graphene nanoribbons from first principles. Applied Physics Letters, 94(22), 223105.

[3] Qin, Z., Qin, G., Shao, B., and Zuo, X. (2017). Unconventional magnetic anisotropy in one-dimensional Rashba system realized by adsorbing Gd atom on zigzag graphene nanoribbons. Nanoscale, 9(32), 11657-11666.

[4] Krychowski, D., Kaczkowski, J., and Lipinski, S. (2014). Kondo effect of a cobalt adatom on a zigzag graphene nanoribbon. Physical Review B, 89(3), 035424.

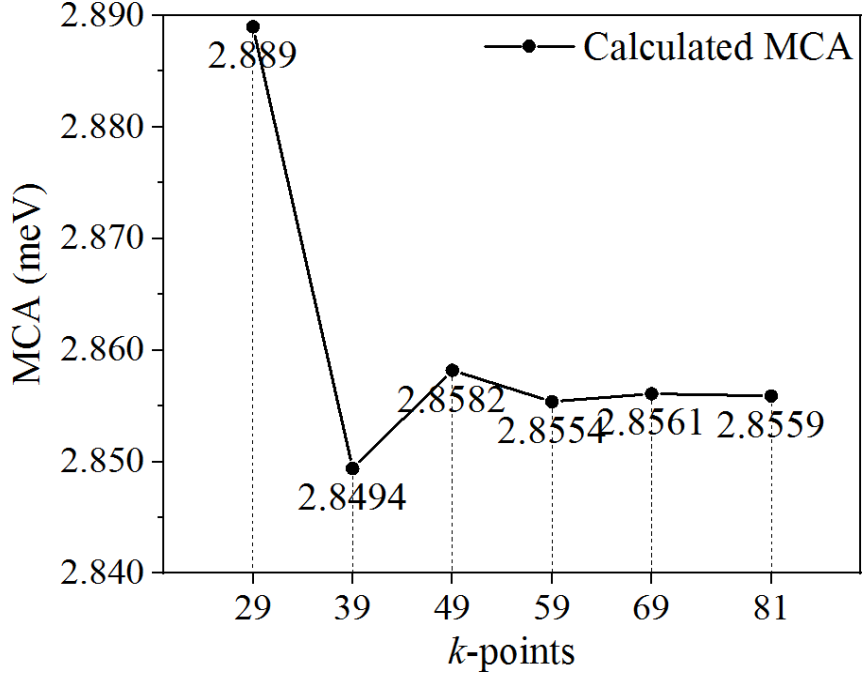

FIG. S1: The influence of the *k*-points along the one-dimensional FBZ on MCA for AFM state of 2Gd-4ZGNR system.

TABLE S I: The fitted values of MCA of 2Gd-4ZGNR for the AFM and FM states based on polynomial fitting formula ($K_0 + K_1 \cos\theta + K_2 \sin^2\theta$), which are corresponding to the fitted curves in FIG. 4 of text. The $K_0$, $K_1$, $K_2$ and the respective standard errors are presented.

| State | $E - E_{0°} = K_0 + K_1 \cos\theta + K_2 \sin^2\theta$ | Standard error |
|---|---|---|
| AFM | $K_0 = 6.740 \times 10^{-3}$ | $A = 0.002$ |
|  | $K_1 = 5.393 \times 10^{-4}$ | $B_1 = 0.002$ |
|  | $K_2 = 2.855$ | $B_2 = 0.003$ |
| FM | $K_0 = 5.590 \times 10^{-3}$ | $A = 0.009$ |
|  | $K_1 = 4.266 \times 10^{-4}$ | $B_1 = 0.007$ |
|  | $K_2 = 4.458$ | $B_2 = 0.014$ |

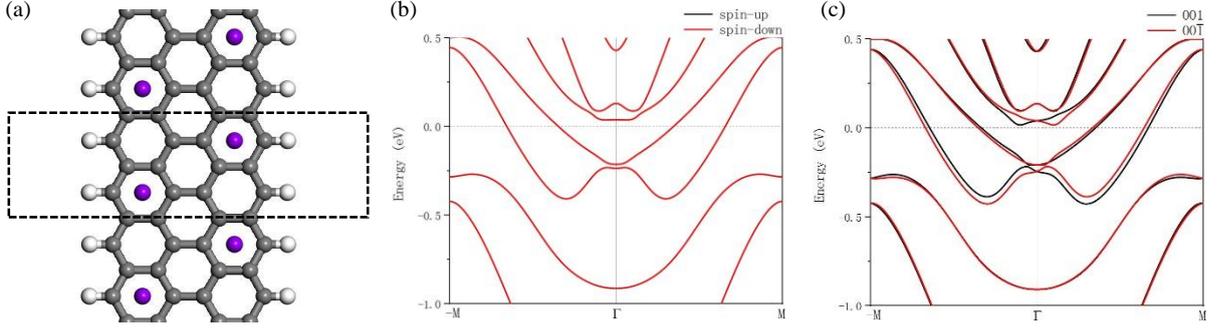

FIG. S2: (a) The top view of the equilibrium structure for 2Gd-4ZGNR structure corresponding to a coverage of two Gd adatoms per 16 carbon atoms. The unit cell of structure is indicated by broken lines. Gray, purple and white balls represent carbon, gadolinium and hydrogen atoms, respectively. (b) The band dispersion relations of non-SOC perturbation for AFM state. (c) The band structures of SOC-perturbed with 001 and 00$\bar{1}$ magnetization directions for AFM state. Note that the fermi level is set to zero.

TABLE S II: Matrix elements of angular momentum operators $L_z$, $L_y$ and $L_x$ with the cubic harmonics of $d-$orbitals.

| $\|u\rangle$ $\langle o\|$ | $L_z$ | | | | | $L_y$ | | | | | $L_x$ | | | | |
|---|---|---|---|---|---|---|---|---|---|---|---|---|---|---|---|
| | $d_{xy}$ | $d_{yz}$ | $d_{z^2}$ | $d_{xz}$ | $d_{x^2-y^2}$ | $d_{xy}$ | $d_{yz}$ | $d_{z^2}$ | $d_{xz}$ | $d_{x^2-y^2}$ | $d_{xy}$ | $d_{yz}$ | $d_{z^2}$ | $d_{xz}$ | $d_{x^2-y^2}$ |
| $d_{xy}$ | 0 | 0 | 0 | 0 | $2i$ | 0 | $i$ | 0 | 0 | 0 | 0 | 0 | 0 | $-i$ | 0 |
| $d_{yz}$ | 0 | 0 | 0 | $i$ | 0 | $-i$ | 0 | 0 | 0 | 0 | 0 | 0 | $-\sqrt{3}i$ | 0 | $-i$ |
| $d_{z^2}$ | 0 | 0 | 0 | 0 | 0 | 0 | 0 | 0 | $-\sqrt{3}i$ | 0 | 0 | $\sqrt{3}i$ | 0 | 0 | 0 |
| $d_{xz}$ | 0 | $-i$ | 0 | 0 | 0 | 0 | 0 | $\sqrt{3}i$ | 0 | $-i$ | $i$ | 0 | 0 | 0 | 0 |
| $d_{x^2-y^2}$ | $-2i$ | 0 | 0 | 0 | 0 | 0 | 0 | 0 | $i$ | 0 | 0 | $i$ | 0 | 0 | 0 |

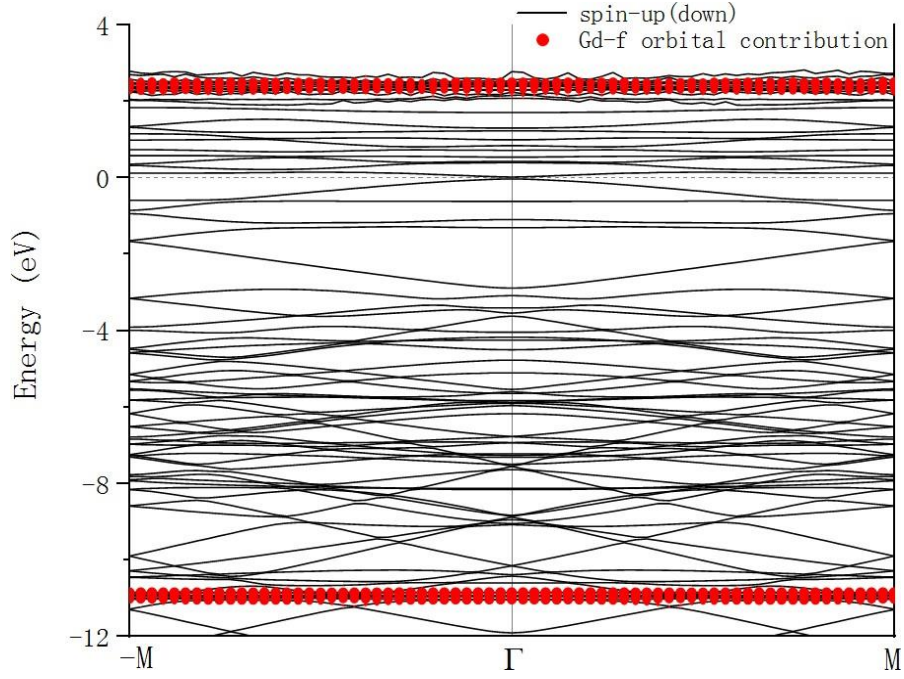

FIG. S3: The spin-up (spin-down) band structures with the distribution of Gd $4f$-orbital states.

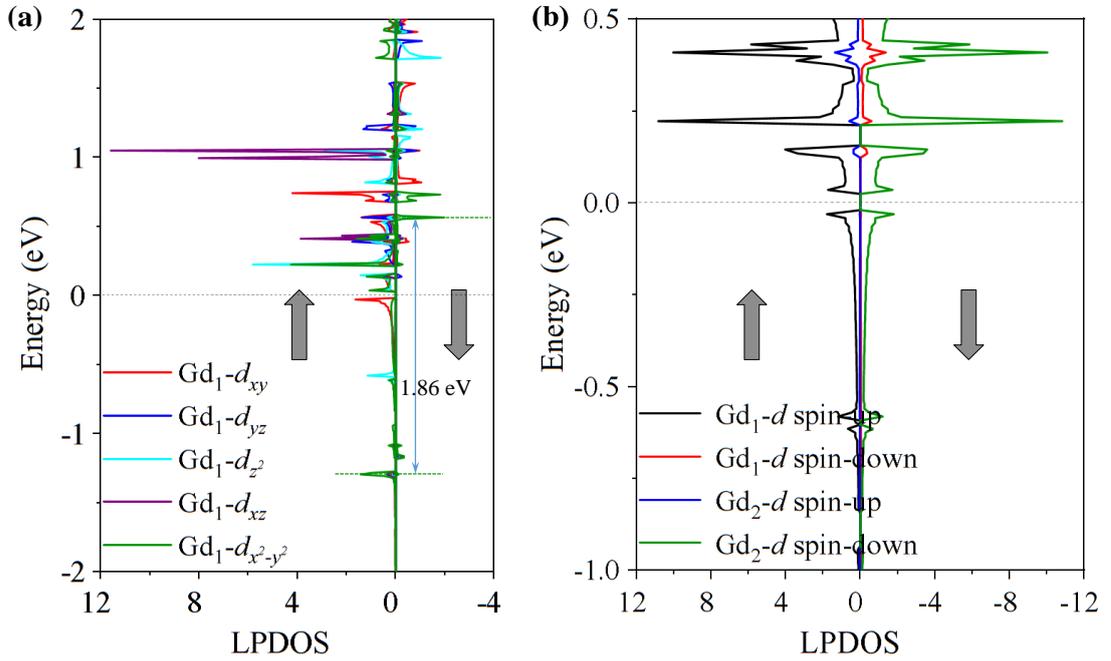

FIG. S4: For AFM state. (a) The local projected density of states (LPDOS) about $5d-$orbital-resolved contributions of $Gd_1$ adatom. (b) The spin-up/down LPDOS of $5d-$orbital of $Gd_1$ and $Gd_2$ adatoms.